\documentclass{elsart3}

 \usepackage{graphicx}

\usepackage{amssymb}

\begin{document}

\begin{frontmatter}



\title{The Dark Side of the Universe}


\author{Katherine Freese}

\address{Michigan Center for Theoretical Physics, Physics Dept. University
of  Michigan, Ann Arbor, MI}

\begin{abstract}
I will begin by reviewing the evidence for Dark Matter in the
Universe, as well as the candidates for dark matter.  At most 20\% of
the dark matter in galaxies can be in the form of MACHOs (Massive
Compact Halo Objects); the remainder appears to be some unknown exotic
component.  The most sensible candidates from the point of view of
particle physics are axions and WIMPs (Weakly Interacting Massive
Particles), where WIMPs may be supersymmetric particles.  Three recent
claims of possible detection of WIMP dark matter are tantalizing and will be
discussed: the DAMA annual modulation, the HEAT positron excess, and
gamma-rays from the Galactic Center.  In addition, I will discuss the
dependence of signals in detectors on the mass distribution in the
Galactic Halo.  In particular, the Sagittarius stream can be a smoking
gun for WIMP detection.

\end{abstract}

\end{frontmatter}

\section{The Dark Matter Problem}
\label{sec:outline}
The universe consists of 4\% ordinary atoms (baryonic matter) 
$\sim 26$\% Dark Matter, and $\sim 70$\% Dark
Energy.  Measurements of the Cosmic Microwave Background Radiation
(CMBR) have determined the total mass density and the geometry of the
universe \cite{wmap}.  The fact that ordinary atoms are 4\% is inferred from
element abundances in Big Bang Nucleosynthesis as well as from the
CMBR.  
The 'best' matter density has been determined by WMAP \cite{wmap} to
be $\Omega_m h^2 = 0.135 (+0.008,-0.009)$ where the best fit value of
the Hubble constant is $h=0.7$. The components are $\Omega_\nu h^2 <
0.0076$ (95\% C.L.), $\Omega_b h^2 = 0.0224 \pm 0.0009$, and
$\Omega_{DM} h^2 = 0.113 (+0.008,-0.009)$ in units of 1.879$\times
10^{-29}$ g/cm$^3$ where subscript $DM$ refers to dark matter.

Evidence for the 70\% dark energy in the universe comes from
observations of distant supernovae \cite{sn}: the supernovae are dimmer than
expected, as is most easily explained by an accelerating universe.
There are two different approaches to the dark energy: (i) a vacuum
energy such as a cosmological constant or time-dependent vacuum may be
responsible \cite{fafm}, or (ii) it is possible that General Relativity is
incomplete and that Einstein's equations need to be modified
\cite{modgenrel}.  Note, however, that this dark energy does not
resolve or contribute to the question of dark matter in galaxies,
which remains as puzzling (if not more) than twenty years ago.

95\% of the mass in galaxies and clusters of galaxies is made of an
unknown dark matter component.  This fact is known from (i) rotation
curves, (ii) gravitational lensing, and (iii) hot gas in clusters.
The speeds of objects in orbit around the centers of galaxies are
determined by the mass interior to the radius of the orbit.  Contrary
to what would be expected from the luminous mass in galaxies, these
rotation curves are flat to very large radii, as can only be explained
if there is an order of magnitude more dark than luminous
matter. Lensing provides another way to detect dark matter: it makes
light bend.  Telescopes monitor the light from distant objects
such as galaxies or quasars: the amount of intervening mass determines
how much this light is bent.  Lensing effects can be responsible for
multiple images of the source object, can cause the source to appear
brighter (microlensing), or can distort the shape of the source
(e.g. shear).  
Recent data from the SLOAN Digital Sky Survey, e.g., have found that
our Milky Way Galaxy is five times more massive than previously
thought and extends out to almost a Mpc in radius \cite{SLOAN}.

Another piece of evidence for dark matter comes from the hot gas in
clusters.  X-ray images of the Coma Cluster, a rich cluster of
galaxies, taken by the ROSAT satellite \cite{COMA} demonstrate that
COMA has a large amount of hot gas. Without a significant dark matter
component in the cluster to provide a gravitational potential well,
the hot gas would evaporate and we wouldn't see it. The majority of
the mass in galaxies and clusters clearly consists of a dark matter
component.

\section{Dark Matter Candidates}
\label{sec:candidates}

There is a plethora of dark matter candidates. The most simple are
MACHOs, or Massive Compact Halo Objects, as these would be made of
ordinary matter in the form of fainst stars or stellar remnants.
However, there are not enough of these to completely resolve the
question.  Of the nonbaryonic candidates, the most popular are the
WIMPS (Weakly Interacting Massive Particles) and the axions, as these
particles have been proposed for other reasons in particle physics.
Ordinary massive neutrinos are too light to be cosmologically
significant, though sterile neutrinos remain a possibility.  Other
candidates include primordial black holes, nonthermal WIMPzillas, and
Kaluza-Klein particles which arise in higher dimensional theories.

\subsection{MACHOs}

MACHO candidates include faint stars, planetary objects (brown
dwarfs), and stellar remnants (white dwarfs, neutron stars, and black
holes).  Microlensing experiments (the MACHO \cite{alcock} and EROS
\cite{eros} experiments) as well as a combination of other
observational (HST) and theoretical results \cite{graff} have shown
that MACHOs less massive than 0.1 solar masses are insignificant in
the Galaxy.  However, there is a detection \cite{alcock} of a roughly 20\% halo
fraction made of $\sim 0.5 M_\odot$ objects which might be made of
stellar remnants such as white dwarfs.  We 
found a number of constraints: the progenitors produce observable
element abundances (C,N,He), they require an enormous mass budget, the
initial mass function must be extremely sharply peaked, and, most
important, the progenitors produce observable infrared radiation. Our
conclusion from these constraints is that at most 20\% of the Galactic
Halo can be made of stellar remnants \cite{ffgwp}.

An interesting recent development \cite{bgs} is a candidate MACHO
microlensing event that has been found in M87, a giant elliptical
galaxy in the VIRGO cluster that is 14 Mpc away. This candidate, which
must be confirmed in future HST observations with more statistics, is
consistent with the results of the MACHO data discussed above.

\begin{figure}
 \includegraphics[width=0.5\textwidth]{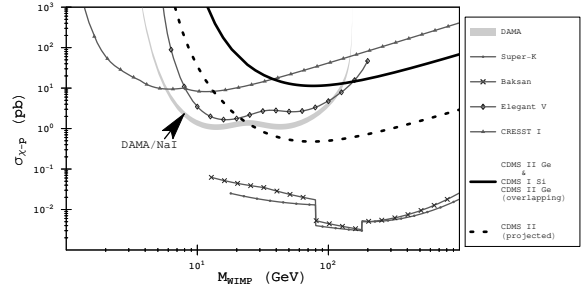}
 \caption{
   WIMP-proton cross-section limits for the case $a_n = 0$.
   Super-K and Baksan rule out the DAMA results over their analysis
   ranges and CRESST I limits DAMA at low masses, but WIMPs between
   6-13 GeV are consistent with all results for this case.
   }
 \label{fig:protonCS}
\end{figure}

\subsection{Axions}
\label{sec:axions}

The good news is that cosmologists don't need to ``invent'' new
particles.  Two candidates already exist in particle physics for other
reasons: axions and WIMPs.  Axions with masses in the range
$10^{-(3-6)}$ eV arise in the Peccei-Quinn solution to the strong-CP
problem in the theory of strong interactions.  Axion bounds
\cite{rosenberg} from the ADMX cavity experiment are approaching the
remaining parameter range.  I wanted also to mention an idea we
recently proposed for inflation with the QCD axion \cite{fls}.  In chain
inflation, the potential looks like a tilted cosine, and the universe
tunnels from higher to lower minima in stages, with a fraction of an
efold at each stage, adding to sufficient inflation.

\section{WIMPs (Weakly Interacting Particles)}
\label{sec:WIMPs}

WIMPs are also natural dark matter candidates from particle phyiscs.
The relic density of these particles comes out to be the right
value: $\Omega_\chi h^2 = (3 \times 10^{-26} {\rm cm}^3/{\rm sec})
/ \sigma_{ann}$, where the annihilation cross section $\sigma_{ann}$
of weak interaction strength automatically gives the right answer. 
The best WIMP candidate is motivated by Supersymmetry (SUSY):
the lightest neutralino in the Minimal Supersymmetric Standard Model.
Supersymmetry in particle theory is designed to keep particle
masses at the right value.  As a consequence, each particle we
know has a partner: the photino is the partner of the photon,
the squark is the quark's partner, and the selectron is the
partner of the electron.  The lightest superysmmetric partner is
a good dark matter candidate (see the reviews by \cite{jkg}).

There are four ways to search for dark WIMPs.  In direct detection
experiments, the WIMP scatters off of a nucleus in the detector, and a
number of experimental signatures of the interaction can be detected
\cite{gw,dfs}.  In indirect detection experiments, neutrinos are
detected from the Sun or Earth that arise as annihilation products of
captured WIMPs; the first papers suggesting this idea were \cite{SOS}
in the Sun and \cite{fkw} in the Earth.  A third way to detect WIMPs
is to look for anomalous cosmic rays from the Galactic Halo: WIMPs in
the Halo can annihilate with one another to give rise to antiprotons,
positrons, or neutrinos \cite{ellis}.  Fourth, neutrinos,
Gamma-rays, and radio waves may be detected as WIMP annihilation
products from the Galactic Center \cite{gonsilk}.

In direct detection experiments, the event rate (number of events/(kg of 
detector)/(keV of recoil energy) is
\begin{equation}
\label{eq:dRdE}
dR/dE 
= {\rho \sigma_0 F^2(q) \over 2 m \mu^2}
\int_{v>\sqrt{ME/2\mu^2}} {f(v,t) \over v} d^3v
\end{equation}
$f(v,t)$ is the WIMP velocity
distribution, $\mu$ is the reduced mass of the WIMP/nucleus system,
$F(q)$ is the nuclear form factor, and the cross section is
\begin{equation}
\label{spinindep}
\sigma_0 = {A^2 \mu^2 \over \mu_p^2} \sigma_p \,\,\,\,\,\,\, 
{\rm spin -independent;}
\end{equation}
\begin{equation}
\label{spindep}
\sigma_0 = {4 \mu^2 \over \pi} |\langle S_p \rangle G_p
+ \langle S_n \rangle G_n |^2 \,\,\,\,\,\, {\rm spin-dependent}
\end{equation}
where the spin-independent cross section scales as atomic
number squared $(A^2)$ and the spin-dependent cross section
depends on the spin content of the nucleon.

{\it {\bf Three Claims of possible WIMP dark matter detection:}}
In the past few years there have been three claims of possible WIMP dark matter
detection: (1) the DAMA annual modulation, (2) the HEAT postiron
excess, and (3) Gamma-rays from the Galactic Center.  

(1) DAMA: In 1986, I was part of a collaboration \cite{dfs} that
suggested using the annual modulation of a WIMP signal to
differentiate it from background (see also \cite{ffg}).  Because the
Sun orbits around the Galactic Center, we are moving into a wind of
WIMPs.  Since the Earth also travels around the Sun, the relative
velocity varies with the time of year, giving
rise to a modulation in the count rate.  The DAMA experiment
\cite{DAMA} has seven years of data with exactly this type of annual
modulation at the 6$\sigma$ level. However, a WIMP interpretation is
controversial.  In fact, a spin-independent cross section with
canonical Maxwellian halo is excluded by the null results of the
CDMS-II experiment \cite{CDMS} (but see also \cite{gongel}).  We asked
whether WIMPs with spin-dependent cross sections explain
both the positive DAMA results as well as the null results from other
experiments \cite{sgf}.  While we found that neutron-only interactions
are ruled out, proton-only or mixed interactions can still survive as
explanations of DAMA data for WIMP masses in the 6-13 GeV range.  The
most stringent bounds arise from SuperKamiokande.  We note that if the
WIMPs are not neutralinos, and are not their only antiparticles, then
the SuperK results do not apply and the WIMP mass could still be as
large as 100 GeV and be in agreement with DAMA and all other existing
data. Figure 1 illustrates our results.

(2) Gamma rays: WIMP annihilation at the Galactic Center?  Recently there have
been claims from both the CANGAROO \cite{CANGAROO} and HESS
\cite{HESS} experiments of detections of Gamma Rays from the Galactic
Center, each with a very different energy spectrum. One can fit the
CANGAROO data with a 2 TeV WIMP, and the HESS data with a 20 TeV WIMP
\cite{horns}.  However, it is not easy to get this intensity in SUSY
models, and it may be easier to explain the data with an astrophysical
origin rather than a SUSY detection.  
Also, Finkbeiner \cite{fink} suggested that excess
microwave emission observed by WMAP in the inner Galaxy (inner $\sim
1-2$kpc) may be interpreted as dark matter annihilation in the inner
galaxy.  Currently 
alternative astrophysical explanations exist for the photons
from the Galactic Center.

In the future, a signal more compelling than the gamma-ray continuum
would be a gamma-ray line, which is characteristic of WIMP
annihilation \cite{bergstrom}.  GLAST may observe such a line below 80
GeV.

(3) Positron excess: The HEAT balloon, using two entirely different
instruments, found an anomaly in the cosmic ray positron flux
\cite{HEAT}.  One possible explanation is dark matter annihilation, as
studied by \cite{begf}; a boost factor of at
least 30 would be required for a thermal SUSY explanation.  A second possible
explanation is that we do not understand cosmic ray propagation.


There have been three recent claims of possible WIMP detection:
DAMA annual modulation, Gamma-rays from the Galactic Center, and the
HEAT positron excess.  All three of these signals are very tantalizing
hints of dark matter detection, but all three may have more
conventional explanations.  Currently the identity of the dark matter
remains an enigma.

\section{Shape of Galactic Halo}

The dark matter distribution in the Halo of the Galaxy affects the
signal in detectors, as seen in Eq.(\ref{eq:dRdE}).  Numerical
simulations indicate that galaxies formed by mergers of smaller
objects.  Recently, one such stream has been found: the Sagittarius
stream.  On the other side of the Galactic Center is a small dwarf
galaxy, Sagittarius, which is being shredded up by our Milky Way: this
is a merger in progress.  Two tidal streams of stars are seen coming
off of the Sagittarius Galaxy.  One of them is seen to pass near the
Solar System.  These streams are likely to contain dark matter, in an
abundance that adds an amount (1-20)\% of the local halo density.  
We calculated the effect of this dark matter stream in detectors
\cite{fgnl}.  This contribution enhances the count rate in dark matter
detectors, but only up to a cutoff in energy in the energy recoil
spectrum. We found that the location of the cutoff changes with the
time of year, so that there is an annual modulation both of the rate
and of the endpoint energy as shown in Figure 3.  The stream also
shifts the peak date of the annual modulation of the signal from June
2 to another date, depending on the stream density.  With a
directional detector such as DRIFT-II \cite{DRIFT}, the stream would
stick out like a sore thumb. In sum, the Sagittarius stream increases
the count rate in detectors up a cutoff in the energy spectrum, the
cutoff location moves in time, the stream sticks out like a sore thumb
in directional detectors, it changes the date of the peak in the
annual modulatoin, and the stream provides a smoking gun for WIMP
detection.

 \begin{figure}
\includegraphics[width=0.45\textwidth]{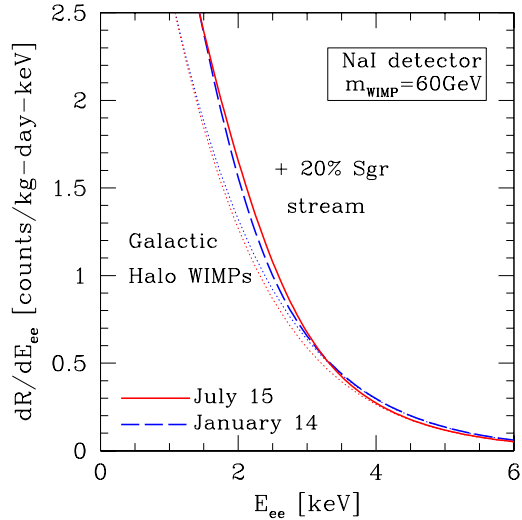}
\caption{Count rate of 60 GeV WIMPs vs.  recoil energy.  The dotted
  lines show the count rate from Galactic (isothermal) halo WIMPs
  alone.  The solid and dashed lines show the step in the count rate
  (in July and January) if we include the Sgr stream WIMPS.  In this
  plot, the stream contributes an additional 20\% to the local Halo
  density. }

\end{figure}

\section{Conclusion}
\label{sec:conclusion}
I have discussed the evidence for dark matter in the universe:
rotation curves, lensing, hot gas in clusters, and the cosmic
microwave background radiation.  Future goals would be to find both
the baryonic and non-baryonic contributions to dark matter.  MACHOs
can provide at most 20\% of the dark matter in galaxies.  The
remainder appears to be some form of exotica. The most sensible
candidates from the point of view of particle physics are axions and
WIMPs, where WIMPs may be supersymmetric particles.  Recently there
have been three claims of possible WIMP detections, and the theoretical
implications of each was discussed.  These three claims are currently
tantalizing but not confirmed.  The expectations for signals in
detectors depend on Halo models, and I showed the implications of the
Sagittarius stream in detectors.



\end{document}